\documentclass{article}
\usepackage{spconf,amsmath,graphicx,url}
\usepackage{multirow}
\usepackage[
separate-uncertainty = true,
multi-part-units = repeat
]{siunitx}
\usepackage{stfloats}
\usepackage{booktabs}
\usepackage{color,soul}
\usepackage[table,xcdraw]{xcolor}
\usepackage{multicol}

\title{Learning utterance-level representations through token-level acoustic latents prediction for Expressive Speech Synthesis}

\name{
	\begin{tabular}{c}
		Karolos Nikitaras$^{\star}$,
		Konstantinos Klapsas$^{\star}$,
		Nikolaos Ellinas$^{\star}$,
		Georgia Maniati$^{\star}$,\\
		June Sig Sung$^{\dagger}$,
		Inchul Hwang$^{\dagger}$, 
		Spyros Raptis$^{\star}$,
		Aimilios Chalamandaris$^{\star}$,
		Pirros Tsiakoulis$^{\star}$
	\end{tabular}
}

\address{$^{\star}$ Innoetics, Samsung Electronics, Greece \\
	$^{\dagger}$ Mobile eXperience Business, Samsung Electronics, Republic of Korea}

\makeindex

\begin{document}
	\ninept
	\maketitle
	\begin{abstract}

	This paper proposes an Expressive Speech Synthesis model that utilizes token-level latent prosodic variables in order to capture and control utterance-level attributes, such as character acting voice and speaking style.
	Current works aim to explicitly factorize such fine-grained and utterance-level speech attributes into different representations extracted by modules that operate in the corresponding level.
	We show that the fine-grained latent space also captures coarse-grained information, which is more evident as the dimension of latent space increases in order to capture diverse prosodic representations.
	Therefore, a trade-off arises between the diversity of the token-level and utterance-level representations and their disentanglement.
	We alleviate this issue by first capturing rich speech attributes into a token-level latent space and then, separately train a prior network that given the input text, learns utterance-level representations in order to predict the phoneme-level, posterior latents extracted during the previous step.
	Both qualitative and quantitative evaluations are used to demonstrate the effectiveness of the proposed approach.
	Audio samples are available in our demo page. \footnote{https://innoetics.github.io/publications/utt-repr-latent/index.html}

	\end{abstract}

	\section{Introduction}
	\label{sec:intro}

	Advances in text-to-speech (TTS) neural systems have resulted in models that are capable of synthesizing high quality speech.
	There is a plethora of models and variations that can be broadly categorized 
	and grouped in various ways. Some categorizations of the TTS models are as follows:
	a) autoregressive (AR) models \cite{shen2018natural, shen2020non} 
	versus parallel or non-autoregressive (NAR) models \cite{ren2020fastspeech, donahue2020end, kim2021conditional}, b) attention based models \cite{shen2018natural} versus duration-informed ones \cite{ren2020fastspeech, shen2020non, donahue2020end}, and c) end-to-end models \cite{kim2021conditional, donahue2020end} versus chained models (acoustic model followed by a vocoder model)  \cite{shen2018natural,shen2020non,ren2020fastspeech}.
	Although, such models have achieved very high quality synthetic speech,
	there are still very important problems that need to addressed,
	such as the control-ability of TTS models, the disentanglement of various speech factors (e.g. speaker identity, recording conditions, speaking style, emotion), expressive and emotional speech generation, etc.
	
	\cite{hsu2018hierarchical, hsu2019disentangling} factorize and control different attributes such as speaker identity and acoustic conditions.
	This work focuses on prosody modeling, where efforts have been made to model and control such attributes by performing at different scales.
	\cite{skerry2018towards, wang2018style, zhang2019learning, shechtman2019sequence} capture the salient features of the utterance by learning coarse-grained representations.
	There are many works that model the prosody in a fine-grained manner \cite{lee2019robust, sun2020fully, vioni2021prosodic, klapsas2021word, ren2020fastspeech}.
	Among them, explicit prosodic features are used in \cite{shechtman2019sequence, vioni2021prosodic, ren2020fastspeech}, while the rest learn latent representations in the corresponding resolution.

	There are also works that perform hierarchical prosody modeling \cite{elias2021parallel, nakata2022predicting, bae2022hierarchical}.
	\cite{elias2021parallel} incorporates the VAE framework \cite{kingma2014adam} by conditioning the phoneme-level representations on utterance-level ones, utilizing a speaker-specific prior.
	\cite{nakata2022predicting} proposes a single-speaker, audiobook speech synthesis model based on Fastspeech 2 \cite{ren2020fastspeech}.
	They leverage character acting voice annotations and capture the acting style variation, in utterance-level, using the VQVAE framework \cite{van2017neural}.
	\cite{raitio2022hierarchical} is also a Fastspeech2-based system that utilizes explicit utterance-wise prosody attributes.

	In this work, we demonstrate that current hierarchical approaches face a trade-off between the diversity of different resolution latent representations and their disentanglement.
	We show that the phoneme-level latent space also captures utterance-level information, which is more evident as the dimension of latent space increases in order to capture diverse prosodic representations.
	We demonstrate that this fact has a negative impact on Prior Network's generalization performance, which is typically responsible for predicting finer-level latents, given text and coarser-level representations, that are utilized during inference time.
	We address this issue by proposing a two-stage training scheme of a multi-scale latent variable system.
	We first capture multi-resolution speech attributes within a phoneme-level latent space and then, separately train a Prior Network that predicts those representations while learning coarse-grained acoustic information.
	
	All in all, the contributions of this work are the following:
	
	\begin{itemize}
			\item we illustrate the diversity-disentanglement trade-off of current hierarchical latent variable TTS systems
			\item we address this trade-off by proposing an alternative approach
			\item we provide a detailed evaluation process by using several objective metrics on both posterior and prior sampling as a function of the number of latent dimensions
	\end{itemize}

	The rest of the paper is organized as follows.
	In Section \ref{ssec:model-architecture}, we present the proposed system and its main components, namely the FVAE \ref{ssec:FVAE} and the Prior Network \ref{ssec:Prior-Network}. 
	The performance evaluation of those modules follows in sections \ref{section:reconstruction-performance} and \ref{section:prior-network-performance}, respectively.
	All alternative models are described, evaluated and compared to the proposed system in section \ref{section:baselines}.
	Finally, section \ref{section:conclusions} concludes our work.

%

\begin{figure*}[t]
	\begin{minipage}[b]{.5\linewidth}
		\centering
		\centerline{\includegraphics[width=.8\linewidth]{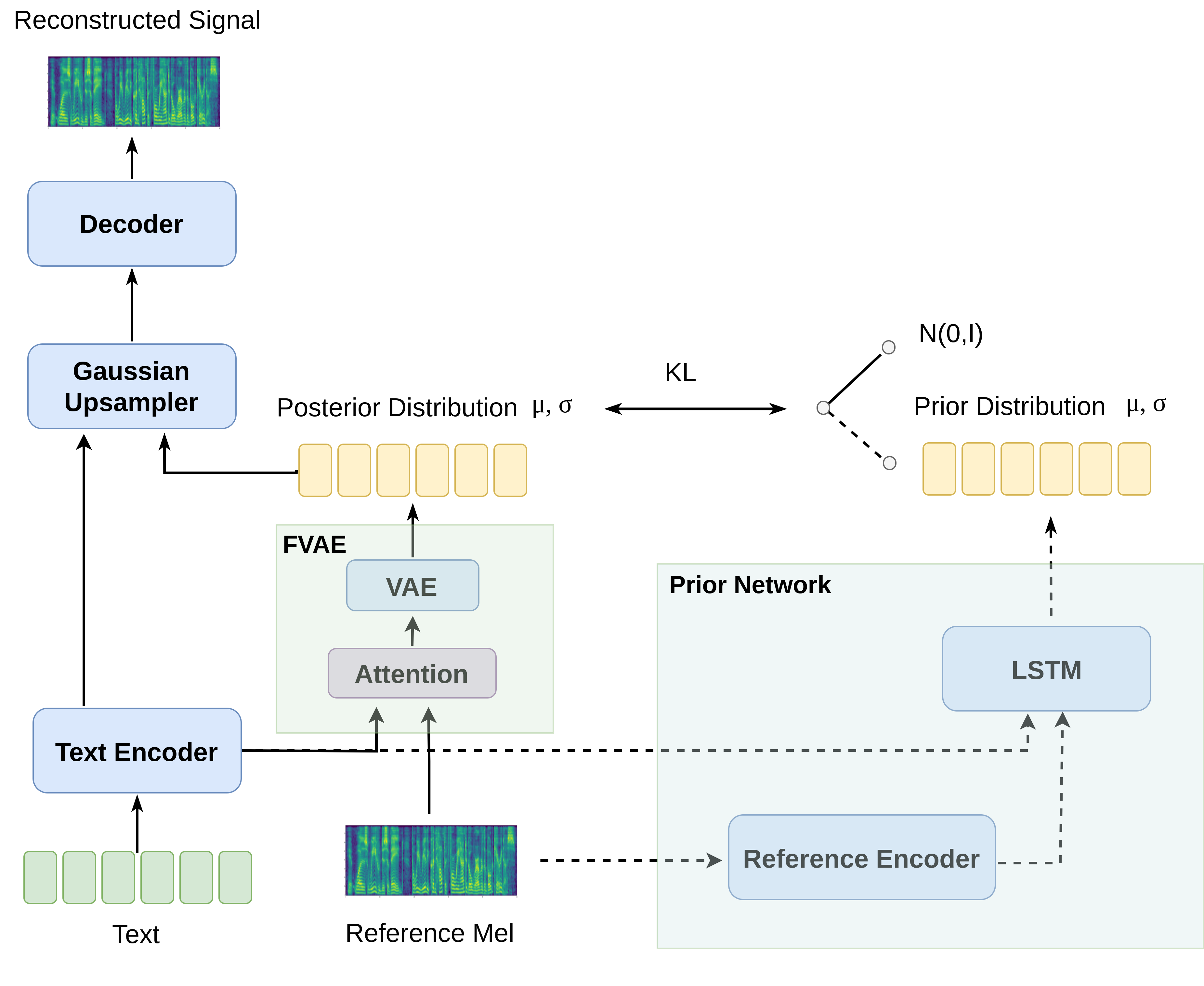}}
		\caption{Model architecture of the proposed approach. Dashed lines correspond to the second training stage where Prior Network is trained.}
	\end{minipage}
	\label{fig:architecture}
	\begin{minipage}[b]{.5\linewidth}
		\centering
		\centerline{\includegraphics[width=.8\linewidth]{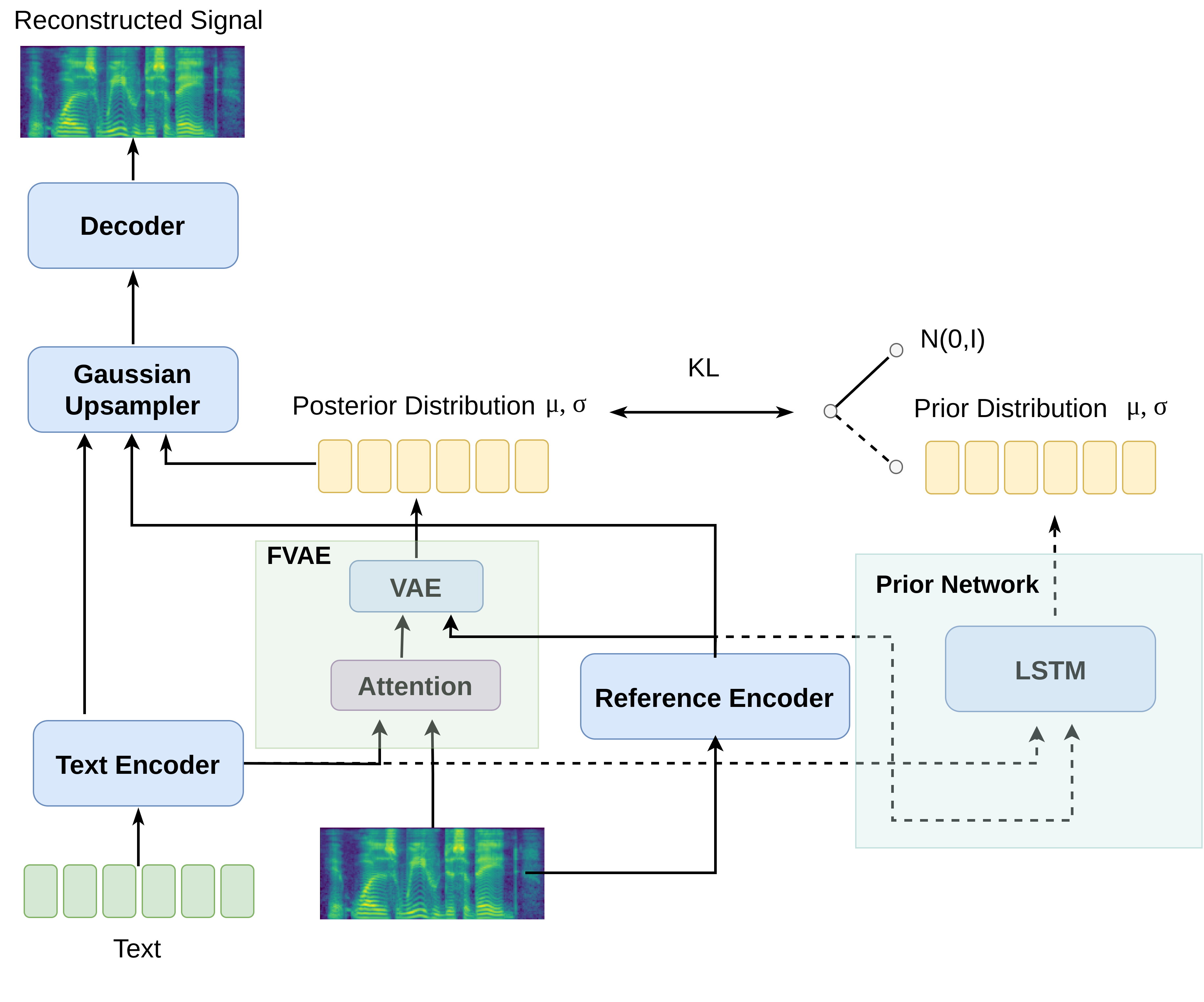}}
		\caption{The overall architecture of a baseline hierarchical system.}
	\end{minipage}
	\label{fig:current_arch}
	\vspace{-10pt}
\end{figure*}

	
	\newpage 
	\section{Model Architecture}
	\label{ssec:model-architecture}
	
	Our system, shown in Figure~\ref{fig:architecture}, is based on the Non-Attentive Tacotron \cite{shen2020non} architecture.
	It passes the input ids $X= (x_1,\dots,x_N)$ of length N to the Text Encoder, which outputs the hidden sequence $H= (h_1,\dots,h_N)$ of the same length.
	Attention mechanism is used to align the target spectrogram sequence $Y= (y_1,\dots,y_T)$ with the sequence H.
	The resulting sequence is mapped into the latent representations $P= (p_1,\dots,p_N)$ using the VAE framework.
	Per-token duration information $D= (d_1,\dots,d_N)$ is being used in order to upsample the phone-wise concatenated sequences H and P.
	$U= (u_1,\dots,u_T)$ is the upsampled sequence, where T= $\sum_{i=1}^{N} d_{i}$ is the total number of mel-spectrogram frames.
	The autoregressive (AR) Decoder Network predicts the acoustic features $Y'= (y'_1,\dots,y'_T)$ given the U sequence, one frame at a time.
	During training, teacher forcing is being used.
	
	During the second stage of training, we train the Prior Network.
	It takes in the text hidden sequence H and the target spectrogram Y.
	By passing Y through the Reference Encoder, the utterance-level representation g is being extracted.
	We broadcast g and concatenate it with the H sequence phone-wise, and feed the resulting sequence to the Gaussian AR prior that outputs the prior sequence $P'= (p'_1,\dots,p'_N)$.
	Prior Network is trained by minimizing the Kullback-Leibler (KL) divergence between the posterior and prior distributions, P and P', respectively.
	Prior Network additionally performs duration prediction, by utilizing an extra dimension whose target is the ground-truth duration information.
	
	Furthermore, the acoustic features used by the model are the 20 bark-scale cepstral coefficients plus pitch period and pitch correlation matching the corresponding the LPCNet-based vocoder \cite{valin2019lpcnet}.
	On the other hand, FVAE and Reference Encoder both take as input sequences of 80-dim mel-scale filterbank frames, as we observed in preliminary experiments that they produced better results compared to the baseline LPCNet-based features used by the vocoder.
	Input text is first normalized and converted into a phoneme sequence by a traditional TTS front-end module.

	\subsection{FVAE}
	\label{ssec:FVAE}
	
	In the first stage of training, we train the FVAE \cite{sun2020fully}.
	The target spectrogram is first aligned with the token encodings using location-sensitive attention \cite{chorowski2015attention}, following \cite{sun2020fully}, where token encodings are used as the queries, while spectrogram frames as both the keys and the values.
	This way, each phoneme is expressed as a weighted sum of the raw spectrogram frames.
	The VAE framework \cite{kingma2013auto} is being used in order to map the aligned, token-level sequence into continuous, latent, acoustic features.
	Unlike \cite{sun2020fully}, scheduled training and conditional dependency are not used.
	The latent acoustic features are concatenated with the phoneme encodings and passed to the Decoder Network.

	\subsection{Prior Network}
	\label{ssec:Prior-Network}
	
	During the second stage of training, we train the Prior Network which consists of the Reference Encoder and the Gaussian AR prior.
	
	\subsubsection{Reference Encoder}
	Reference Encoder contains 3 convolutional blocks, followed by a bi-LSTM and a mean pooling layer.
	It maps each utterance into a single vector that gets concatenated with the phoneme encodings and passed to the the following module. 
	
	\subsubsection{Gaussian AR Prior}
	We use a single-layer LSTM, like \cite{sun2020generating}, to model the temporal coherency in the continuous, token-level, latent space that resulted by the FVAE training during the previous stage \ref{ssec:FVAE}.
	The output at each step is a diagonal Gaussian distribution whose mean and standard deviation depend on the latent features of all the previous tokens.
	We use an additional dimension in order to predict the per-token duration by also modeling the temporal dependency between the duration and the rest of the acoustic attributes.
	
	
	$$
	p\left(\mathbf{z'}_n \mid \mathbf{k}_{<n}\right)=\mathcal{N}\left(\mathbf{z'}_n ; \boldsymbol{\mu}\left(\mathbf{k}_{<n}\right), \boldsymbol{\sigma}\left(\mathbf{k}_{<n}\right)\right)
	$$
	
	where
	
	\begin{itemize}
	
	\item $\mathbf{z'}_{n}:\{\mathbf{z}_{n}, \mathbf{D}_{n}\}$
		
	\item $\mathbf{k}_{<n}:\{\mathbf{z'}_{<n}, \mathbf{H}_{<n},\mathbf{g}\}$
	
	\end{itemize}
	

	Prior Network is trained by minimizing the KL between the posterior and the prior distributions.

	\section{Experiments}
	\label{section:experiments}
	
	The proposed model is evaluated on the Blizzard Challenge 2013 single-speaker audiobook dataset \cite{prahallad2013blizzard}.
	This dataset contains 147 hours US English speech with highly varying prosody, recorded by a female professional speaker.
	
	To quantify the reconstruction performance, we measure the mel-cepstral distortion (MCD) \cite{kubichek1993mel} and the F0 Frame Error (FFE) \cite{chu2009reducing}.
	MCD evaluates the timbral distortion, while FFE evaluates the reconstruction of the F0 track.
	The rest of the metrics are used to evaluate both posterior and prior sampling.
	Like \cite{sun2020generating}, we measure the variance of prosody factors such as energy, F0 and duration at the token level, aiming to quantify the prosodic variance that our system achieves.
	For the purpose of this paper we refer to the different acting voices (characters) as different speakers, even though they are performed by the same voice actor.
	We measure the d-vector similarity between the target and the predicted signals, in order to estimate how effectively our model preserves the speaker timbre.
	D-vectors are utterance-level vectors trained on the speaker verification task using the generalized end-to-end (GE2E) loss \cite{wan2018generalized}.
	We use a system trained on VoxCeleb2 \cite{Chung18b}.
	We also report the word error rate (WER) from an ASR model trained on LibriSpeech \cite{panayotov2015librispeech}.
	Finally, we conduct subjective listening tests to measure the naturalness and the style similarity.
	
	A good system should be able to generate natural audio samples with high prosodic diversity that matches the utterance-level characteristics of the reference utterance.
	We encourage readers to listen to the audio examples on the accompanying web page \footnote{https://innoetics.github.io/publications/utt-repr-latent/index.html}.
	
%
%
%
	
	\begin{table}[h]
		\centering
		\caption{Reconstruction performance of the proposed approach. "D-vec sim" refers to the d-vector similarity.}
		\label{tab:reconstruction-performance-table}
		\begin{tabular}{cccccc}
			\hline
			&                              &                               & \multicolumn{2}{c}{Prosody stddev}                           &              \\
			model                              & MCD                          & FFE                           & E                            & F0                            & d-vec sim \\ \hline
			Real Speech                        & -                            & -                             & 5.25                         & 44.62                         & -            \\ \hline
			\cellcolor[HTML]{FFFFFF}nat        & \cellcolor[HTML]{FFFFFF}5.84 & \cellcolor[HTML]{FFFFFF}29.48 & \cellcolor[HTML]{FFFFFF}4.64 & \cellcolor[HTML]{FFFFFF}23.44 & 0.844        \\ \hline
			\cellcolor[HTML]{FFFFFF}2-dim hvae & \cellcolor[HTML]{FFFFFF}5.11 & \cellcolor[HTML]{FFFFFF}22.74 & \cellcolor[HTML]{FFFFFF}4.87 & \cellcolor[HTML]{FFFFFF}27.94 & 0.864        \\
			8-dim mvae                         & 3.85                         & 5.97                          & 5.07                         & 41.32                         & 0.935        \\
			16-dim mvae                        & 3.31                         & 4.64                          & 5.14                         & 42.91                         & 0.954        \\
			32-dim mvae                        & 3.02                         & $\textbf{4.44}$                          & 5.27                         & 42.63                         & 0.963        \\
			64-dim mvae                        & $\textbf{2.93}$                         & 4.45                          & $\textbf{5.27}$                         & $\textbf{43.54}$                         & $\textbf{0.966}$        \\ \hline
		\end{tabular}
	\end{table}
	
	\subsection{FVAE Posterior Sampling}
	\label{section:reconstruction-performance}
	
	Table~\ref{tab:reconstruction-performance-table} shows the reconstruction, or copy synthesis, performance regarding the first training stage of the proposed system.
	As a baseline, we implement a Non-Attentive Tacotron model which does not utilize a reference encoder. 
	We also train several FVAE models using a variable number of latent dimensions.
	We also train several FVAE models using a variable number of latent dimensions.
	Models using a conditional dependency across the latent dimensions, as in \cite{sun2020fully}, are denoted as hierarchical (\text{hvae}), whereas when this dependency is not used we simply refer to the model as multi-level (\text{mvae}).
	Using the ground-truth duration labels, we perform posterior sampling on 1000 utterances (real speech).
	We also provide some prosody related metrics that estimate the variance of the phoneme-level Energy and F0.
	
	As the number of the latent dimensions increases, both FFE and MCD metrics get lower.
	Similarly, the variance of the prosody-related metrics gets improved and approach the real speech levels.
	FVAE better preserves the speaker timbre, which is an utterance-level speech attribute, when using more dimensions.
	We encourage the readers to listen to the audio examples and observe the significant improvement in terms of all these aspects of reconstruction performance as a function of the number of the latent dimensions.

	%
	
	
	\subsection{Prior Network Performance}
	\label{section:prior-network-performance}
	
	During the second stage, we train the Prior Network which outputs a prior distribution trained to be close to the posterior distribution that resulted in the previous stage.
	In ~\ref{section:reconstruction-performance} we showed that there are utterance-level factors, such as speaker timbre, that have been captured within the token-level latent space.
	Linguistic content itself is not informative about such variations.
	Therefore, we employ utterance-level acoustic representations extracted by the Reference Encoder.
	
	\vspace{0cm}
	\begin{table}[ht]
		\centering
		\caption{Prior sampling performance of the proposed approach. "CG dims" refers to coarse-grained dimensions.}
		\label{tab:prior-sampling-table}
		\begin{tabular}{ccccccc}
			\hline
			&                     &      &              & \multicolumn{3}{c}{Prosody stddev} \\
			model                        & CG dims      & WER  & d-vec sim & F0         & E         & Dur       \\ \hline
			Real Speech                  & -                   & 4.5  & -            & -          & -         & -         \\ \hline
			nat                          & 0                   & 5.8  & 0.75         & 18.26      & 3.49      & 18.54     \\ \hline
			8-dim mvae                   & \multirow{4}{*}{16} & $\textbf{5.39}$ & 0.813        & 35.83      & 4.00      & 28.33     \\
			16-dim mvae                  &                     & 5.81 & 0.826        & $\textbf{41.76}$      & $\textbf{4.18}$      & $\textbf{32.27}$     \\
			32-dim mvae                  &                     & 5.89 & $\textbf{0.831}$        & 35.56      & 4.14      & 31.63     \\
			64-dim mvae                  &                     & 5.72 & 0.813        & 34.47      & 3.68      & 31.24     \\ \hline
			\multirow{4}{*}{64-dim mvae} & 8                   & 5.74 & 0.806        & 32.05      & 3.80      & 24.94     \\
			& 16                  & $\textbf{5.72}$ & 0.813        & 34.47      & 3.68      & 31.24     \\
			& 32                  & 6.02 & 0.827        & $\textbf{37.53}$      & $\textbf{3.99}$      & $\textbf{33.27}$     \\
			& 64                  & 6.10 & $\textbf{0.835}$        & 35.24      & 3.75      & 32.03     \\ \hline
		\end{tabular}
	\end{table}

	Table~\ref{tab:prior-sampling-table} evaluates the prior sampling performance.
	We use 15 unseen, reference utterances that represent a wide range of utterance-level variations such as acting voices, speech rate and loudness, as well as acoustic conditions.
	For each target signal, we synthesize 50 out-of-domain sentences.
	We use several combinations between the pre-trained FVAE latent space and the trainable, utterance-level one.
	Also, we synthesize these sentences using a regular Non-Attentive model, which makes no use of a reference signal.
	
	All 4 models that learn a 16-dim utterance-level latent space, achieve comparable performance in terms of speech intelligibility and prosody modeling.
	The same holds for speaker similarity, even though Table~\ref{tab:reconstruction-performance-table} illustrated that higher-dimensional FVAE latent spaces better preserve the speaker timbre.
	This suggests that it is a matter of capacity of the coarse-grained representations learned by the Prior Network for the prediction of the posterior latents.
	In the 4 following models we increased the number of the utterance-level dimensions, only to observe that more dimensions are associated with better speaker similarity at a cost in speech intelligibility.
	Also, except the 8-dim case, the rest 3 models perform comparably in terms of prosody modeling.

		\begin{table*}[ht]
		\centering
		\caption{Posterior and Prior sampling performance of baseline systems}
		\label{tab:prior-posterior-sampling-baseline-table}
		\begin{tabular}{ccccccc|ccccc}
			\hline
			&                & \multicolumn{5}{c|}{Posterior Sampling}                                                                         & \multicolumn{5}{c}{Prior Sampling}                    \\
			&                &                              &                               &           & \multicolumn{2}{c|}{Prosody stddev}  &      &           & \multicolumn{3}{c}{Prosody stddev} \\ \hline
			model       &  CG dims & MCD                          & FFE                           & d-vec sim & F0                            & E    & WER  & d-vec sim & F0         & E         & Dur       \\ \hline
			Real Speech & -              & -                            & -                             & -         & 44.62                         & 5.25 & 4.5  & -         & -          & -         & -         \\ \hline
			nat         & 0              & \cellcolor[HTML]{FFFFFF}5.84 & \cellcolor[HTML]{FFFFFF}29.48 & 0.844     & \cellcolor[HTML]{FFFFFF}23.44 & 4.64 & 5.8  & 0.75      & 18.26      & 3.49      & 18.54     \\ \hline
			nat         & 16             & 5.33                         & 25.07                         & 0.891     & 26.70                         & 4.97 & 5.78 & 0.835     & 32.91      & $\textbf{4.13}$      & 29.57     \\ \hline
			hvae 2      &             & 4.74                         & 21.78                         & 0.912     & 25.57                         & 5.19 & $\textbf{5.71}$ & $\textbf{0.842}$     & 29.57      & 3.77      & 25.08     \\
			mvae 8      & 16               & 3.68                         & 5.67                          & 0.945     & 41.80                         & $\textbf{5.27}$ & 6.02 & 0.838     & $\textbf{33.78}$      & 4.03      & 25.88     \\
			mvae 64     &                & $\textbf{2.98}$                         & $\textbf{4.41}$                          & $\textbf{0.965}$     & $\textbf{45.06}$                         & 5.06 & 7.76 & 0.809     & 29.08      & 3.95      & $\textbf{34.85}$     \\ \hline
		\end{tabular}
	\end{table*}
	
	\subsection{Alternative Approaches and Performance Evaluation}
	\label{section:baselines}

	In this section, we compare to several baseline models, all based on Non-Attentive (NAT), by performing both objective and subjective evaluation.
	We consider a simple NAT model for comparison purposes, as well as another NAT with its Decoder Network conditioned on global representations learned by a Reference Encoder of the same architecture as ours.
	The following 3 models represent the commonly used hierarchical approach, shown in Figure~\ref{fig:current_arch}, where finer-level representations are conditioned on coarser-grained ones.
	In particular, we condition FVAE models on utterance-level representations extracted by a Reference Encoder.
	Note that during inference time, the NAT version that uses 16-dimensional global representations depends on a reference signal, which is being used as an extra input both at the Decoder Network and the Duration Predictor.
	FVAE systems additionally condition the FVAE latents on the target signal and during inference time, depend on a Prior Network similar to ours.
	
	\vspace{-.1cm}
	\subsubsection{Objective Evaluation}
	\label{section:baselines-objective-evaluation}
	
	The evaluation process of posterior and prior sampling is similar to those of the sections ~\ref{section:reconstruction-performance} and ~\ref{section:prior-network-performance}, respectively.
	As shown at Table~\ref{tab:prior-posterior-sampling-baseline-table}, NAT has a poor performance in prosody modeling.
	Across the rest of the models, which all use the target spectrogram as a reference signal, FVAE beats NAT in terms of reconstruction performance, while their prior sampling performance is comparable.
	It is worth noting that the more FVAE latent dimensions that are being used, the better the speaker similarity is within the reconstructed signals.
	However, this does not hold regarding the prior sampling.
	Both facts indicate that such variation has been captured within the token-level posterior latents, rather than the coarse-grained representations and this difference in performance grows as the number of dimensions increases.
	To discourage FVAE latents from capturing utterance-level speech attributes, one could lower their dimensionality at a cost of prosody modeling performance.
	Thus, such approach involves the trade-off between the disentanglement of the different resolution latent spaces and their diversity.
	Finally, there is higher variance in prosody related attributes of FVAE systems that use more dimensions.
	
	\vspace{-.1cm}
	\subsubsection{Subjective Evaluation}
	
	We conducted subjective listening tests with native English speakers asked to rate both the naturalness of speech samples and the style similarity of the synthesized speech compared to the reference signal.
	Regarding the Naturalness MOS, we randomly selected 10 references utterances each used to synthesize 15 out-of-domain sentences.
	In terms of the Similarity MOS, we manually selected 15 reference utterances associated with a wide variety of utterance-level speech attributes, such as acting voice, speech pace, speech loudness, as well as acoustic conditions.
	For each target utterance we synthesized 5 unseen, in-domain sentences which correspond to utterances that are mapped into points, within the coarse-grained latent space, that are nearby to that of the target signal.
	The latter utterances correspond to the Ground-truth signals in the similarity test, while arbitrary utterances to the Ground-truth signals in the Naturalness test.		
	Table~\ref{tab:mos-table} shows the MOS results, where Ground-truth signals achieved the higher scores in both tests.
	All systems except the last one, correspond to baseline models described in section~\ref{section:baselines}.
	Recall that the regular NAT system makes no use of reference signal and so, we just synthesized the corresponding sentences.

	\begin{table}[htb]
	\centering
	\caption{Similarity and Naturalness MOS results. FG and CG refer to fine-grained and coarse-grained dimensions, respectively.}
	\label{tab:mos-table}
	\begin{tabular}{ccccc}
		\hline
		model        & FG dims 			 & CG dims       & \begin{tabular}[c]{@{}c@{}}Similarity\\ MOS\end{tabular} & \begin{tabular}[c]{@{}c@{}}Naturalness \\ MOS\end{tabular}        \\ \hline
		Grount-truth & -                 & -                    & $4.6 {\scriptstyle \pm 0.03}$                         & $4.6 {\scriptstyle \pm  0.04}$           \\ \hline
		nat          & 0                 & 0                    & $2.93 {\scriptstyle \pm 0.08}$ 						& $3.59 {\scriptstyle \pm 0.08}$          \\ \hline
		nat          & 0                 & 16                   & $3.43 {\scriptstyle \pm 0.1}$                          & $\textbf{3.38} {\scriptstyle \pm 0.06}$ \\ \hline
		hvae         & 2                 &                      & $3.25 {\scriptstyle \pm 0.11}$                         & $3.37 {\scriptstyle \pm 0.06}$          \\
		mvae         & 8                 &                      & $3.35 {\scriptstyle \pm 0.11}$                        & $3.29 {\scriptstyle \pm 0.06}$          \\
		mvae         & 64                & \multirow{-3}{*}{16} & $2.91 {\scriptstyle \pm 0.11}$                         & $2.94 {\scriptstyle \pm 0.06}$           \\ \hline
		proposed     & 16                & 16                   & $\textbf{3.45} {\scriptstyle \pm 0.1}$                 & $3.35 {\scriptstyle \pm 0.06}$         \\ \hline
	\end{tabular}
\end{table}

	In terms of the similarity test, the proposed systems seems to beat all the alternative systems.
	The regular NAT system totally fails in preserving the coarse-grained characteristics of the reference utterance.
	Right above, the NAT version that only learns utterance-level characteristics performed comparably to the proposed system.
	The next 3 models that condition finer-level representations to coarser-level ones, achieve significantly lower performance.
	It is worth mentioning that the baseline system that uses the largest FVAE latent dimensions performs worst.
	This verifies the trade-off that we mention in section~\ref{section:baselines-objective-evaluation}.
	
	Regarding the naturalness MOS, the regular NAT system outperforms all the alternative systems.
	However, it should be noted that NAT benefits from learning an averaged speaking style, which is unaffected by reference utterances during inference. 
	That might also explain the performance degradation across all the rest of the systems.
	The MOS scores of the NAT version, that captures coarse-grained information, are comparable with the proposed system.
	All the rest of the NAT versions that incorporate the FVAE module perform the worst.
	
	\section{Conclusions}
	\label{section:conclusions}
	
	In this work we demonstrate a trade-off between the diversity of different resolution latent representations and their disentanglement and address it by proposing a two-stage training scheme.
	We first capture multi-resolution speech attributes within a phoneme-level latent space and then, separately train a Prior Network that predicts those representations while learning coarse-grained acoustic information.
	Our approach better preserves the coarse-grained attributes of the reference utterances, while maintaining comparable performance in terms of speech intelligibility.
	
	
	\bibliographystyle{IEEEbib}
	\bibliography{refs}

	%


\end{document}